\documentclass[fleqn,12pt,twoside]{article}
\usepackage{espcrc1}

\usepackage{graphicx}

\newcommand{\AmS}{{\protect\the\textfont2
  A\kern-.1667em\lower.5ex\hbox{M}\kern-.125emS}}

\newcommand{\sfrac}[2]{{\textstyle\frac{#1}{#2}}}
\newcommand{\Lagr}{\mathcal{L}}

\title{$SU(3)$ Chiral Effective Field Theories\\
                --- A Status Report ---}

\author{B. Borasoy\address{Physik Department, Technische Universit\"at M\"unchen, \\ 
        85747 Garching, Germany}%
        \thanks{Financial support of the Deutsche Forschungsgemeinschaft is gratefully acknowledged.}}
       
\begin{document}

\maketitle

\begin{abstract}
A personal overview on the present status of $SU(3)$ chiral perturbation theory
in the baryonic sector is given. Recent developments are presented
and remaining challenges are pointed out.
\end{abstract}

\section{Introduction}

Within an effective field theory heavy particles are frozen and reduced to static sources,
whereas active, light particles are treated as dynamical degrees of freedom.
In order to describe the dynamics, an effective Lagrangian is constructed which incorporates all relevant
symmetries and symmetry breaking patterns of the underlying fundamental theory.

For the strong interactions the corresponding fundamental theory is Quantum Chromodynamics
which has quark and gluon fields as explicit degrees of freedom in the Lagrangian.
Due to confinement, however, neither quarks nor gluons are observed as free
particles in nature; they rather combine to colorless objects, to hadrons and possibly glueballs.
A closer look at the hadronic spectrum reveals a characteristic gap which separates
the eight lightest pseudoscalar mesons -- the three pions, the four kaons
and the eta --  from the other hadrons, such as the vector mesons or the nucleons.
This pattern can be explained as spontaneous breaking of $SU(3)_L \times SU(3)_R$
chiral symmetry of the QCD Lagrangian with massless quarks which leads to eight
Goldstone bosons. Due to the actual finite size of the quark masses the (pseudo-)Goldstone
bosons acquire small masses and are identified with the octet of pseudoscalar mesons,
$(\pi, K, \eta)$.

Hadron physics at low energies is governed by the eight Goldstone bosons and can be
described in an efficient way by the effective field theory of QCD,
chiral perturbation theory (ChPT). ChPT provides a model-independent framework
with the same symmetries and symmetry breaking patterns as QCD.
Systematically developed first in the mesonic sector by Gasser and Leutwyler \cite{GL},
it can be extended to include baryons, see {\it e.g.} \cite{BSW,GSS,JM}.
In this overview, we will focus on applications of baryon ChPT for three light flavors
with a particular emphasis on recent developments and remaining challenges in this field.

In the next section general construction principles for the effective Lagrangian
are outlined. The octet baryon masses and $\sigma$ terms of the nucleon are discussed in
Section~3. Results of axial vector couplings as well as ordinary nonleptonic and radiative 
weak hyperon decays are presented in Secs.~4 to 6.

\section{Chiral effective Lagrangian}

In this section, some guiding principles will be presented for the construction of the 
effective Lagrangian.
As QCD is invariant under  chiral $SU(3)_L \times SU(3)_R$ transformations
in the limit of massless up, down and strange quarks, the effective Lagrangian ${\cal L}$
decomposes in the following way
\begin{equation}
{\cal L} = {\cal L}_0 + {\cal L}_{\mbox{\scriptsize{sb}}}  
\end{equation}
with ${\cal L}_0 $ being invariant under chiral $SU(3)_L \times SU(3)_R$ transformations.
The second term incorporates chiral symmetry breaking 
patterns due to non-zero quark masses $(m_u, m_d, m_s)$ and vanishes in the chiral limit
of massless quarks. The quark masses are assumed to be small so that the 
explicitly chiral symmetry breaking terms in ${\cal L}_{\mbox{\scriptsize{sb}}}$ 
can be treated perturbatively.

Let us first restrict ourselves to the purely mesonic sector.
The eight Goldstone bosons are most conveniently summarized in a matrix
$U(x) \in SU(3)$ with
\begin{equation}
U(x) = u^2(x) =\exp \left(i \sfrac{\sqrt{2}}{f} \, \phi (x) \right) ,
\end{equation}
where $f$  is the pseudoscalar decay constant in the chiral limit and $\phi$ is given by
\begin{equation}
 \phi =
\left(\begin{array}{ccc} \frac{1}{\sqrt{2}} \pi^0 + \frac{1}{\sqrt{6}} \eta &\phantom{+} \pi^+  
& \phantom{+} K^+ \\[8pt]
\phantom{+}  \pi^- &  -\frac{1}{\sqrt{2}} \pi^0 + \frac{1}{\sqrt{6}} \eta & \phantom{+} K^0\\[8pt]
\phantom{+} K^- &  \phantom{+} \bar{K}^0 & \phantom{+} - \frac{2}{\sqrt{6}} \eta  \end{array}\right) .
\end{equation}
The effective Lagrangian is a function of $U, \partial_\mu U$ and the
quark mass matrix ${\cal M}= \mbox{diag} (m_u, m_d, m_s)$,
$\; \Lagr = \Lagr ( U, \partial U, \partial^2 U, \ldots , {\cal M})$,
and is expanded in powers of ${\cal M}$ and $\partial_\mu U$ which amounts to an expansion
in powers of meson momenta and masses.
For $\Lagr_0$ we obtain the chiral expansion
\begin{equation}
\Lagr_0 = \Lagr_0^{(2)} + \Lagr_0^{(4)} + \ldots 
\end{equation}
with the superscript $(i)$ denoting the $i^{\mbox{\scriptsize{th}}}$ chiral order.
As the Lagrangian is a Lorentz scalar, only even chiral powers are allowed.
The leading term $\Lagr_0^{(2)}$ (omitting for the moment external vector and 
axial vector fields) reads
\begin{equation}
\Lagr_0^{(2)} = \frac{f^2}{4} \langle \partial_\mu U^\dagger 
\partial^\mu U  \rangle 
= \frac{1}{2}\langle \partial_\mu \phi \partial^\mu \phi  \rangle 
+ \frac{1}{12 f^2}  \langle [\phi,\partial_\mu \phi] [\phi,\partial^\mu \phi]  \rangle  + \ldots \; ,
\end{equation}
where we have expanded $U$ in the meson fields $\phi$.
The first term in the $\phi$ expansion is the kinetic piece of the meson propagators,
while the second term represents a four-meson interaction with a coupling
constant fixed by chiral symmetry. Vertices with higher powers in $\phi$ 
are denoted by the ellipsis and $\langle \ldots  \rangle $ is the trace in flavor space.
At higher chiral orders, however, new additional coupling constants appear that
need to be determined by experiment.

The explicitly chiral symmetry breaking piece ${\cal L}_{\mbox{\scriptsize{sb}}}$
is expanded analogously
\begin{equation}
\Lagr_{\mbox{\scriptsize{sb}}} = \Lagr_{\mbox{\scriptsize{sb}}}^{(2)} 
+ \Lagr_{\mbox{\scriptsize{sb}}}^{(4)} + \ldots 
\end{equation}
with the leading term
\begin{equation}
\Lagr_{\mbox{\scriptsize{sb}}}^{(2)} = B_0 \frac{f^2}{2} \langle 
 {\cal M} (U + U^\dagger)\rangle  .
\end{equation}
The quark mass matrix counts as second chiral order and
always enters in combination with $B_0 = - \langle 0| \bar{q} q| 0 \rangle/f^2$,
the order parameter of spontaneous symmetry violation. Taken by itself, ${\cal M}$
depends on the running scale of QCD, whereas the product $B_0 {\cal M}$
is renormalization group invariant.

The effective field theory can be extended to
include the ground state $SU(3)$ baryon octet $B$ consisting of the nucleons and hyperons
which are collected in a $3 \times 3$ matrix 
\begin{equation}
B =
\left(\begin{array}{ccc} \frac{1}{\sqrt{2}} \Sigma^0 + \frac{1}{\sqrt{6}} \Lambda &\phantom{+} \Sigma^+  
& \phantom{+} p \\[8pt]
\phantom{+}  \Sigma^- &  -\frac{1}{\sqrt{2}} \Sigma^0 + \frac{1}{\sqrt{6}} \Lambda & \phantom{+} n\\[8pt]
\phantom{+} \Xi^- &  \phantom{+} \Xi^0 & \phantom{+} - \frac{2}{\sqrt{6}} \Lambda  \end{array}\right) 
\end{equation}
and at leading order in the baryonic sector the effective Lagrangian reads
\begin{equation} \label{barlagr}
\Lagr_{\phi \mbox{\scriptsize{$B$}}}^{(1)} 
= i \langle \bar{B} \gamma_\mu D^\mu B \rangle -  M_0  \langle \bar{B} B \rangle  
 - \sfrac{i}{2} D \langle \bar{B} \gamma_\mu \gamma_5 \{ u^\mu , B \} \rangle 
- \sfrac{i}{2} F \langle \bar{B} \gamma_\mu \gamma_5 [ u^\mu , B ] \rangle .
\end{equation}
$M_0 $ is the common octet baryon mass in the chiral limit, while
$D$ and $F$ are the axial vector couplings of the nucleons which can be determined
from hyperon beta decays. As we will see in Section~4, a good fit to the decays is given by 
$D=0.80$ and $F=0.46$.

The inclusion of external vector and axial vector fields in the effective field theory,
{\it e.g.}, in order to describe semileptonic baryon decays or processes with photon fields, 
promotes the global chiral symmetry to a local one.
The local nature of chiral symmetry requires the replacement of partial derivatives
$\partial_\mu$ by gauge covariant ones which involve external vector and axial vector fields.
For the baryon derivative this implies the replacement
\begin{equation}
\partial_\mu B \to D_\mu B = \partial_\mu B + [\Gamma_\mu,B]
\end{equation}
with the so-called chiral connection $\Gamma_\mu$
\begin{equation}
\Gamma_\mu= \frac{1}{2} [u^\dagger, \partial_\mu u] - \frac{i}{2} 
                        ( u^\dagger r_\mu u + u l_\mu u^\dagger) 
\end{equation}
and the left- and right-handed combinations of the external fields,
$l_\mu = v_\mu - a_\mu$ and $r_\mu = v_\mu + a_\mu$, respectively.
For meson fields the replacement reads
\begin{equation}
\partial_\mu U \to \nabla_\mu U = \partial_\mu U - i r_\mu U + i U l_\mu .
\end{equation}
with the covariant derivative included in
\begin{equation}
u_\mu = i u^\dagger \nabla_\mu U \, u^\dagger .
\end{equation}
Expanding $u_\mu$ in the meson fields yields
\begin{equation}
u_\mu = - 2 i a_\mu - \frac{\sqrt{2}}{f} \partial_\mu \phi + \ldots \;.
\end{equation}
Hence, the two latter terms in Eq.~\ref{barlagr} describe the couplings
of both the mesons and the external axial vector fields to the baryons at leading order.
From this observation one deduces immediately the generalized Goldberger-Treiman relations,
which read, {\it e.g.}, for the nucleon
\begin{equation}
g_{\pi NN} = \frac{g_A^{pn} M_N}{f_\pi}
\end{equation}
with $g_A^{pn} = D+F=1.26$, $M_N$ the nucleon mass, and we have replaced $f$ by the physical value 
for the pion decay constant, $f_\pi = 92.4$ MeV, which is consistent at leading order.
Inserting the experimental value for the $\pi NN$ coupling constant, $|g_{\pi NN}| \approx 13.0$,
one finds agreement at the 2\% level.

The inclusion of baryons introduces a new scale $M_0$ which is close to the scale of
spontaneous chiral symmetry breaking, $\Lambda_\chi = 4 \pi f_\pi \sim 1.2 \mbox{ GeV}$.
Strictly speaking, the scale $M_0$ spoils the chiral counting scheme, {\it i.e.} higher loops
contribute to lower chiral orders.
However, the power counting can be reestablished by treating the fermions as heavy
sources in the non-relativistic framework of heavy baryon ChPT \cite{JM}.
Alternatively, one can evaluate the loops in the relativistic framework with infrared
regularization by isolating in the loop integrals the infrared singularities due to the Goldstone bosons
\cite{ET,BL}.

Another important question is whether the strange quark is light enough to be in the
chiral regime and if it is included appropriately within the standard $SU(3)$ framework of
ChPT. In QCD the values of the current quark masses at $\mu=2$ GeV in the $\overline{\mbox{MS}}$
scheme are $m_u= 1.5-4.5$ MeV, $m_d= 5-8.5$ MeV and $m_s= 80-155$ MeV \cite{PDG}.
Thus the masses of the up and down quark, $m_u, m_d$, are light compared to any
hadronic scale, {\it e.g.}, $m_u, m_d \ll \Lambda_{QCD} \sim 200 $ MeV.
The mass of the strange quark, on the other hand, is comparable in size,
$m_s \sim \Lambda_{QCD}$. It is therefore legitimate to question whether
$SU(3)_L \times SU(3)_R$ chiral symmetry is a realistic starting point
for a systematic expansion in symmetry breaking interactions.
At the level of the effective theory, the relevant expansion parameter is given
by the ratio $m_K/\Lambda_\chi \sim 0.4$. One may prefer to consider the kaon as a heavy particle
and treat it as a heavy source as, {\it e.g.}, done in heavy kaon ChPT \cite{Ro}. 
Here we follow the standard scenario of $SU(3)$ ChPT in which the strange quark is
treated on equal footing as the up and down quarks.

\section{Baryon masses and $\mbox{\boldmath$\sigma$}$ terms}

Having set up the effective Lagrangian, we now proceed by calculating 
quantities in the scalar sector of baryon ChPT: the baryon masses and the 
$\sigma$ terms. At second chiral order
quark masses enter the baryonic Lagrangian
\begin{equation}
\Lagr_{\phi \mbox{\scriptsize{B}}}^{(2)} 
= 4 B_0 b_0 \langle \bar{B}  B \rangle \langle {\cal M} \rangle 
+ 4 B_0 b_D \langle \bar{B} \{ {\cal M} , B\} \rangle 
+ 4 B_0 b_F \langle \bar{B} [ {\cal M} , B] \rangle 
\end{equation}
with unknown parameters $b_0, b_F, b_D$ to be determined from experiment.
Utilizing this Lagrangian, one calculates the mass splittings of the baryon octet
at leading order in symmetry breaking. We work in the isospin limit $m_u = m_d$
so that there are only four different baryon masses, $(M_N, M_\Lambda, M_\Sigma, M_\Xi)$,
and obtain
\begin{eqnarray}
M_N & =& \tilde{M}_0 - 4 m_K^2 b_D
+ 4 (m_K^2 - m_\pi^2) b_F , \nonumber \\
M_\Lambda & =& \tilde{M}_0 - \frac{4}{3} (m_K^2 - m_\pi^2) b_D , \nonumber \\
M_\Sigma & = & \tilde{M}_0 - 4 m_\pi^2 b_D , \nonumber \\
M_\Xi & =& \tilde{M}_0 - 4 m_K^2 b_D- 4 (m_K^2 - m_\pi^2) b_F .
\end{eqnarray}
Note that we have absorbed the constant contributions from the $b_0$ term to the masses
into $\tilde{M}_0$. Thus the four octet baryon masses are effectively represented in 
terms of three parameters which leads to a sum rule, the Gell-Mann--Okubo
mass relation,
\begin{equation}
M_\Sigma - M_N = \frac{1}{2} (M_\Xi - M_N) + \frac{3}{4} (M_\Sigma - M_\Lambda)
\end{equation}
satisfied experimentally to about 3\%.
By going to higher chiral orders one expects to even ameliorate the situation.
In general, the chiral expansion of the masses reads
\begin{equation}
M_B = M_0 + \sum_q b_q m_q  + \sum_q c_q m_q^{3/2} + \sum_q d_q m_q^2 + \ldots
\end{equation}
with the leading non-analytic contribution at third chiral order
and chiral logarithms at fourth chiral order
are not shown explicitly.
A complete  one-loop calculation has been performed in the heavy baryon approach 
to fourth chiral order in \cite{BM}
\begin{eqnarray}
M_N & =& M_0(1 + 0.34 - 0.35 + 0.24) ,  \nonumber \\
M_\Lambda & =& M_0(1 + 0.69 - 0.77 + 0.54) , \nonumber \\
M_\Sigma & = & M_0(1 + 0.81 - 0.70 + 0.44) , \nonumber \\
M_\Xi & =& M_0(1 + 1.10 - 1.16 + 0.78) ,
\end{eqnarray}
where $M_0$ has been factored out and the numbers denote the relative size of the 
contributions at zeroth, second, third and fourth chiral order, respectively.
Due to the proliferation of new unknown coupling constants at fourth chiral order an exact fit
to the baryon octet masses is possible, but no clear statement can be made on the convergence
of the chiral expansions.
In \cite{BM} the loop integrals have been evaluated in dimensional regularization, where
the large non-analytic corrections at third chiral order arise from the integral
\begin{equation} \label{integral}
\int \frac{d^4k}{(2 \pi)^4} \; \frac{k_i k_j}{[k_0 + i \epsilon] \, [k^2 - m^2+ i \epsilon ]}
 = i \delta_{ij} \frac{I(m)}{24 \pi}
\end{equation}
which in dimensional regularization is calculated to be $I_{dim.reg.} (m)= m^3$
and is sizeable for kaon and eta loops.

Such large non-analytic contributions occur, since baryons are treated as point
like particles within the effective theory, although they have a finite size of
about 1 fm. One must therefore suppress the short distance portions of loop integrals
which are not described appropriately by chiral physics.
This can be achieved, {\it e.g.}, by utilizing a simple dipole regulator \cite{DHB}
\begin{equation}
\left( \frac{\Lambda^2}{\Lambda^2 - k^2} \right)^2 ,
\end{equation}
however, the specific choice of the regulator is not important as long as it
maintains the relevant symmetries. The insertion of the dipole regulator in
Eq.~\ref{integral} yields
\begin{equation}
I_\Lambda (m) = \Lambda^4 \frac{2m + \Lambda}{2 (m + \Lambda)^2} 
\end{equation}
and it is verified that the dipole regulator maintains indeed chiral symmetry, as the power
divergences proportional to $\Lambda^3$ and $\Lambda$ can be absorbed into
$M_0$ and $b_0, b_D, b_F$, respectively. 
The value of the cutoff $\Lambda$ must be chosen in such a way that it suppresses the
short distance portion of the integral but keeps the low energy part. Therefore,
the phenomenologically relevant cutoffs are in the range 
$1/\langle r_B \rangle \le \Lambda  \sim \; 300 \; - \; 600  \; \mbox{MeV}$.

In Table~\ref{table:1} the non-analytic contributions from the integral
in Eq.~\ref{integral}  to the baryon masses are given
both for dimensional and cutoff regularization (in units of GeV).
Clearly, utilization of a cutoff ameliorates convergence problems from large loop effects. 

\begin{table}[htb]
\caption{Non-analytic contributions to the baryon masses in GeV.}
\label{table:1}
\newcommand{\m}{\hphantom{$-$}}
\newcommand{\cc}[1]{\multicolumn{1}{c}{#1}}
\renewcommand{\tabcolsep}{2pc} 
\renewcommand{\arraystretch}{1.2} 
\begin{tabular}{@{}cccc}
\hline
          & \quad dim. reg. \quad  & \quad $\Lambda$ = 300 MeV
             \quad  &   \quad $\Lambda$ = 400 MeV  \quad  \\
\hline
$N$                 & -0.31 &0.02 & 0.03 \\
$\Lambda$           & -0.66 &0.03 & 0.06 \\
$\Sigma$            & -0.62 &0.03 & 0.05 \\
$\Xi$               & -1.03 &0.04 & 0.08 \\
\hline
\end{tabular}\\[2pt]
\end{table}

Further information in the scalar sector of baryon ChPT is contained in the $\sigma$ terms
which measure the strength of the matrix elements $\bar{q} q$ in the nucleon
\begin{eqnarray}
\sigma_{\pi N} (t) & = & \frac{1}{2} ( m_u + m_d) \langle p' | \bar{u} u + \bar{d} d | p \rangle , \nonumber \\
\sigma_{K N}^{(1)} (t) & = &\frac{1}{2} ( \hat{m} + m_s) 
\langle p' | \bar{u} u + \bar{s} s | p \rangle , \nonumber  \\
\sigma_{K N}^{(2)} (t) & =& \frac{1}{2} ( \hat{m} + m_s) 
\langle p' | - \bar{u} u + 2 \bar{d} d +\bar{s} s | p \rangle  ,
\end{eqnarray}
with $t \equiv (p' -p)^2 $ the momentum transfer squared and 
$\hat{m} \equiv ( m_u + m_d)/2$.
The empirical value for the pion-nucleon $\sigma$ term can deduced from
an extrapolation of low-energy pion nucleon scattering data and has been determined
in \cite{GLS} to be
\begin{equation} \label{sigma}
\sigma_{\pi N} (0) = (45 \pm 8) \mbox{MeV } .
\end{equation}
Closely related to the $\sigma$ terms is the strangeness fraction $y$ in the nucleon
which is given by the ratio
\begin{equation}
y =  \frac{\textstyle{2 \langle p | \bar{s} s | p \rangle}}{\textstyle{
   \langle p | \bar{u} u + \bar{d} d | p \rangle}} .
\end{equation}
To leading order in the quark masses and using $SU(3)$ baryon wave functions one obtains
\begin{equation}
\sigma_{\pi N} (0) =  \frac{\textstyle{\hat{m}}}{\textstyle{m_s - \hat{m}}} \; 
   \frac{\textstyle{M_\Xi + M_\Sigma - 2 M_N}}{\textstyle{1-y}}
  =  \frac{\textstyle{26 \, \mbox{MeV}}}{\textstyle{1-y}} .
\end{equation}
Using $\sigma_{\pi N} (0) = 45$ MeV this corresponds to a value of $y=0.42$, {\it i.e.}
the nucleon appears to have a large strangeness content.
However, two orders higher in the chiral expansion this problem is resolved
and one gets \cite{BM}
\begin{equation}
\sigma_{\pi N} (0) =  \frac{\textstyle{(36 \pm 7) \mbox{MeV}}}{\textstyle{1-y}} 
\end{equation}
which translates into $y = 0.2 \pm 0.2$. This result is compatible with zero, but has
a tendency for a non-zero admixture of strange quarks in the nucleon.
The calculation has also been performed in the cutoff scheme  with the
results $\sigma_{K N}^{(1)} (0) \; = 380 \pm 50 \; \mbox{MeV}$,
$\sigma_{K N}^{(2)} (0) \; = 250 \pm 40 \; \mbox{MeV}$,
$y \; = 0.25 \pm 0.05 $, and the strangeness contribution to the nucleon mass
$m_s \langle p | \bar{s} s | p \rangle \; = 150 \pm 50 \; \mbox{MeV} $,
while keeping $\sigma_{\pi N} (0) = 45$ MeV fixed \cite{B}.

Lately, new $\pi N$ scattering data from TRIUMF and PSI have  become available
and more recent extractions of $\sigma_{\pi N} (0)$ range from 45 to 80 MeV
corresponding to a strangeness fraction $y$ from 0.2 to 0.5.
More definite statements about these quantities can only be made, once the new dispersion theoretical
analyses currently under development in Karlsruhe and Helsinki are completed.

\section{Axial vector couplings}

The hadronic axial current for the semileptonic decay $B_i \to B_j l \bar{\nu}_l$
is written in the form
\begin{equation}
\langle B_j | A_\mu | B_i \rangle = \bar{u} (p_j) \bigg(  g_1(q^2) \gamma_\mu \gamma_5 
              - \frac{\textstyle{i g_2(q^2)}}{\textstyle{M_i + M_j}} 
              \sigma_{\mu \nu} q^\nu \gamma_5  
    + \frac{\textstyle{ g_3(q^2)}}{\textstyle{M_i + M_j}} q_\mu \gamma_5  \bigg) u (p_i) .
\end{equation}
The axial vector couplings are defined as $g_A \equiv g_1(0) $
and can be expressed at leading chiral order in terms of the two couplings
$D$ and $F$ from $\Lagr_{\phi \mbox{\scriptsize{$B$}}}^{(1)}$.
A good fit to the experimentally measured $g_A$ is obtained with $D= 0.80,\; F=0.46$
and suggests that $SU(3)$ breaking effects in the data are smaller than $ 10 \%$,
see Table~\ref{table:2}.

\begin{table}[htb]
\caption{Tree level fit to axial vector couplings of the baryon octet.
  Compared are the theoretical results with the experimental values.}
\label{table:2}
\newcommand{\m}{\hphantom{$-$}}
\newcommand{\cc}[1]{\multicolumn{1}{c}{#1}}
\renewcommand{\arraystretch}{1.2} 
\begin{tabular}{@{}cccccccc}
\hline
     & $\; g_A^{pn}$   & $\; g_A^{p \Lambda }$   &    $\; g_A^{\Lambda \Sigma^-}$  
     & $\; g_A^{\Xi^0 \Xi^-}$  &  $\; g_A^{\Lambda \Xi^-}$ &  $\; g_A^{n \Sigma^-}$ 
     & $\; g_A^{\Sigma^0 \Xi^-}$  \\
\hline
     & $D+F$ & $- \frac{1}{\sqrt{6}} (D+3F)$ & $\frac{2}{\sqrt{6}} D$ & $D-F$ & $- \frac{1}{\sqrt{6}} (D-3F)$
     & $D-F$ & $\frac{1}{\sqrt{2}} (D+F)$ \\
Th.  & 1.26 & -0.89 & 0.65 & 0.34 & 0.24 & 0.34 & 0.89 \\
Exp. & 1.267 & -0.89 & 0.60 & & 0.30 &0.34 & 0.93 \\     
\hline
\end{tabular}\\[2pt]
\end{table}

One expects to improve the situation by going to higher chiral order, but
the inclusion of the lowest non-analytic contributions from chiral loops
leads to significant $SU(3)$ breaking which is in disagreement with experiment \cite{BSW}.
Again, application of a cutoff regulator brings the chiral corrections under control
and leads to a well-behaved chiral expansion \cite{DHB,B2}.

\section{Nonleptonic weak hyperon decays}

The nonleptonic weak decays are the dominant hadronic decay mode of the hyperons.
There are seven such decays:
$ \Lambda \to \pi^0 n $, $ \Lambda \to \pi^- p$,
$ \Sigma^+ \to \pi^+ n$, $\Sigma^+ \to \pi^0 p$, $\Sigma^- \to \pi^- n$,
$\Xi^0 \to \pi^0 \Lambda $,  and $ \Xi^- \to \pi^- \Lambda$.
The matrix elements of the decays are described in terms of two amplitudes,
a parity-violating $s$ wave, $A$,  and a parity-conserving $p$ wave, $B$,
\begin{equation}
{\cal A} (B \to B' \pi) = \bar{u}_{B'} (p') ( A + B \gamma_5) u_B (p) .
\end{equation}
For more than three decades nonleptonic hyperon decays have been examined
using effective field theories and there remain two important issues on
the theoretical side.

The first one is the $\Delta I = 1/2$ rule which states that the $\Delta I = 3/2$
amplitudes are suppressed with respect to the $\Delta I = 1/2$ 
counterparts by factors of about twenty. This suppression exists also in kaon
nonleptonic decay and despite considerable theoretical efforts still
no simple explanation for its validity is available. Note, however, that there
may be possibly a $\Delta I = 1/2$ rule violation in hypernuclear decay \cite{DFHT}.
For our purposes, it is justified to neglect the $\Delta I = 3/2$
amplitudes. Isospin symmetry of the strong interactions implies then three relations
for both $s$ and $p$ waves, so that we end up with eight independent decay amplitudes.

The second problem is the so-called $s$ and $p$ wave puzzle.
To lowest order the weak chiral effective Lagrangian is given by two contact terms
\begin{equation}
\Lagr_{\mbox{\scriptsize{$W$}}}^{(0)} 
=  d_{\mbox{\scriptsize{$W$}}} \langle \bar{B} \{ u^\dagger \lambda_6 u , B \} \rangle 
+  f_{\mbox{\scriptsize{$W$}}} \langle \bar{B} [ u^\dagger \lambda_6 u , B ] \rangle 
\end{equation}
with the Gell-Mann matrix $\lambda_6$ and two unknown coefficients $d_{\mbox{\scriptsize{$W$}}}$ 
and $f_{\mbox{\scriptsize{$W$}}}$.

The amplitudes are given at leading order by contact interactions for the
$s$ waves, Figure~1a),  and baryon pole diagrams for the $p$ waves, Figs.~1b) and c).
\begin{figure}[htb]
\begin{center}
\includegraphics[width=13cm]{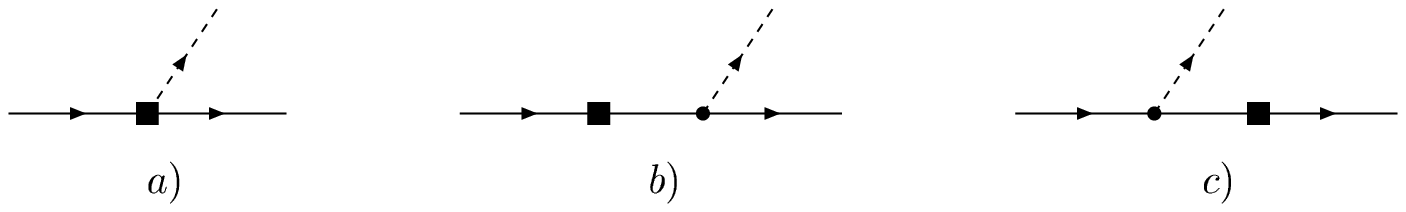}
\end{center}
\caption{Given are the tree diagrams for $s$ waves, (a), and $p$ waves, (b and c).
         The square denotes a weak vertex, while the circle is a strong vertex.
         Baryons and pions are represented by solid and dashed lines, respectively.}
\end{figure}

When trying to fit the data, it turns out that a reasonable simultaneous fit both to
$s$ and $p$ waves is not possible. A good $s$ wave fit can be obtained, but this yields
a poor $p$ wave description. On the other hand, a good $p$ wave representation
yields a poor $s$ wave fit.
In order to overcome this problem, one must go beyond leading order.
In \cite{BSW} the leading non-analytic corrections from the Goldstone boson
loops were computed, but no local counterterms were considered.
The resulting $s$ wave predictions no longer agreed with the data, and the corrections
for the $p$ waves were even larger.
Jenkins reinvestigated this topic by explicitly including spin-3/2 decuplet fields
in the effective theory and was able to restore good agreement between experiment
and theory for the $p$ waves, although the description for the $p$ waves was not satisfactory \cite{Jen}.
A complete one-loop calculation has been performed in \cite{BH} which introduces
more coupling constants than there exist data. An exact fit is possible, but not unique,
even when reducing the number of counterterms by utilizing resonance saturation.
It is unclear what the underlying physics is and the actual values of the 
couplings remain undetermined.

Another intriguing possibility was examined by Le Yaouanc et al., who assert
that a reasonable fit for both $s$ and $p$ waves can be provided by appending
pole contributions from $SU(6) \; (70,1^-)$ states to the $s$ waves \cite{LeY}.
Their calculations were performed in a simple constituent quark model and appear 
to provide a resolution of the $s$ and $p$ wave dilemma.
In \cite{BH2} the validity of this approach has been considered in ChPT.
Adding the contributions from the lowest lying 1/2$^-$ and 1/2$^+$ 
baryon resonant states to the tree level results appears to provide a satisfactory
picture of nonleptonic hyperon decays. The weak contact interactions of these resonances
with the ground state baryon octet are sizeable and parametrize effects from even higher
energies.

Alternatively, the importance of final state interactions has been investigated
within a chiral unitary approach based on coupled channels \cite{BMa}. In this approach
the baryon resonances are generated dynamically instead of being included explicitly and
the phase shifts of $\pi N$ scattering are reproduced at the relevant energies
which guarantees the accurate inclusion of the final state interactions. 
One observes that final state interactions have larger effects than
usually assumed; they are greater than 10\% and should not be omitted.
Although the inclusion of final state interactions improves the overall fit
of $d_{\mbox{\scriptsize{$W$}}}$ and $f_{\mbox{\scriptsize{$W$}}}$ to the data, 
a reasonable description is not obtained. 
This emphasizes again the importance of physics from higher energies which enters
in the effective field theory via higher order weak contact interactions.

\section{Nonleptonic radiative hyperon decays}

The weak radiative hyperon decays
$\Sigma^+ \to p \gamma $, $  \Sigma^0 \to n \gamma$, $ \Lambda \to n \gamma$,
$\Xi^0 \to \Sigma^0 \gamma $, $ \Xi^0 \to \Lambda \gamma$, and $ \Xi^- \to \Sigma^- \gamma $
are described by
\begin{equation}
{\cal A} (B \to B' \gamma) = - \; \frac{\textstyle{i}}{\textstyle{2(M_B + M_{B'})}} 
\; \bar{u}_{B'} (p') \sigma_{\mu \nu}  F^{\mu \nu}  ( C_{\mbox{\scriptsize{$B' B$}}} 
+ D_{\mbox{\scriptsize{$B' B$}}} \gamma_5) u_B (p)
\end{equation}
with $C$ and $D$ being the parity-conserving $M1$ and parity-violating $E1$ amplitudes,
respectively. 
In the $SU(3)$ limit Hara's theorem requires the vanishing of $B_{\mbox{\scriptsize{$B' B$}}}$
for decays between baryon states of a common $U$-spin multiplet (interchange of $s$ and $d$ quark):
$\Sigma^+ \to p \gamma , \quad \Xi^- \to \Sigma^- \gamma$ \cite{Ha}.
In the real world one expects about 20\% $SU(3)$ breaking effects and thus a small
photon asymmetry 
\begin{equation}
\alpha \equiv 
- \frac{\textstyle{2 \mbox{Re} C^* D}}{\textstyle{|C|^2+|D|^2}} .
\end{equation}
But this is in contradiction to the near maximal value of the asymmetry parameter
measured experimentally in polarized $\Sigma^+ \to p \gamma$: 
$\; \alpha_{\mbox{\scriptsize{$\Sigma^+ p$}}}= -0.76 \pm 0.08$ \cite{PDG}.

At lowest order in ChPT only baryon pole diagrams contribute
leading to a vanishing parity-violating amplitude $D$ 
and  asymmetry parameter $\alpha =0$ for all decays.
Recent work involving the calculation of chiral loops has also not led to a resolution,
although slightly larger asymmetries could be accommodated \cite{Neu}.

The inclusion of intermediate $(70,1^-)$ states, on the other hand, has been argued to lead to a resolution
of the asymmetry problem in radiative decay, similar to the case of the $s$ and $p$ wave
puzzle for ordinary hyperon decay \cite{Gav}.
This claim could be confirmed within the chiral framework by including again the 1/2$^-$ and 1/2$^+$ 
baryon resonant states. The same weak contact interactions as in ordinary hyperon
decay contribute and yield relatively large asymmetries \cite{BH3}.

\section{Summary}
We conclude by noting that $SU(3)$ ChPT is an appropriate tool to investigate properties and decays 
of hyperons. Despite recent progress challenging problems remain as illustrated above
such that the study of hyperons continues to be an interesting and active field.

\section*{Acknowledgements}
It is a pleasure to thank the organizers for this pleasant and stimulating conference.

\end{document}